\documentclass[a4paper]{PoS}
\usepackage{amsmath,amssymb,amsfonts}

\newcommand{\calO}{\mathcal{O}}

\title{Ground state charmed meson spectra for $N_f\!=\!2\!+\!1\!+\!1$}

\ShortTitle{Ground state charmed meson spectra for $N_f=2+1+1$}

\author{\speaker{T.~D.~Rae}$^a$, S.~D\"{u}rr$^{a,b}$ \\
        $^a$ Bergische Universit\"{a}t Wuppertal, Germany\\
        $^b$ IAS/JSC Forschungszentrum J\"{u}lich, Germany\\
        E-mail: \email{thrae@uni-wuppertal.de}}

\abstract{
	We present a preliminary study of the charmed meson spectra using the electrically neutral subset of the new Budapest-Marseille-Wuppertal $N_f=2+1+1$ gauge configurations that utilise the 3-HEX smeared clover action. The analysis is performed with a focus on the hyperfine splitting.
	}

\FullConference{The 33rd International Symposium on Lattice Field Theory\\
                14-18 July 2015\\
                Kobe International Conference Center, Kobe, Japan}

\begin{document}

\section{Introduction}

\noindent Experiments such as Babar, Belle, BES III and LHCb continue to provide an abundance of charmed physics results \cite{Brambilla:2010cs, Amhis:2014hma}. Lattice QCD (LQCD) proves to be a vital tool in the interpretation of these results; for a review see \cite{Bouchard:2015pda,El-Khadra:2014sha,Aoki:2013ldr}. For example, LQCD measurements provide access to CKM matrix elements, and thus a check of the SM, from the leptonic and semileptonic decays of $D$ mesons. LQCD could also help to understand the currently unclassified experimental states that cannot be incorporated into the quark model. In addition, the masses and decays of low lying charmonium states have been accurately measured by experiment and provide an excellent test of precision LQCD.  The subject of this proceeding is the hyperfine mass splittings of the doubly charmed $J/\psi$ and $\eta_{cc}$ mesons and the singly charmed $D_s^*$ and $D_s$ mesons. These mass splittings typically show significant cut-off effects for Wilson type fermions (see e.g. \cite{Bali:2011dc} and references therein). It is, therefore, important that this is understood through a careful systematic study of the lattice measurements.

\section{Lattice setup and methodology}
\begin{figure}[h!]
\begin{center}
\includegraphics[width=0.5\textwidth,trim={8 10 5 5}, clip]{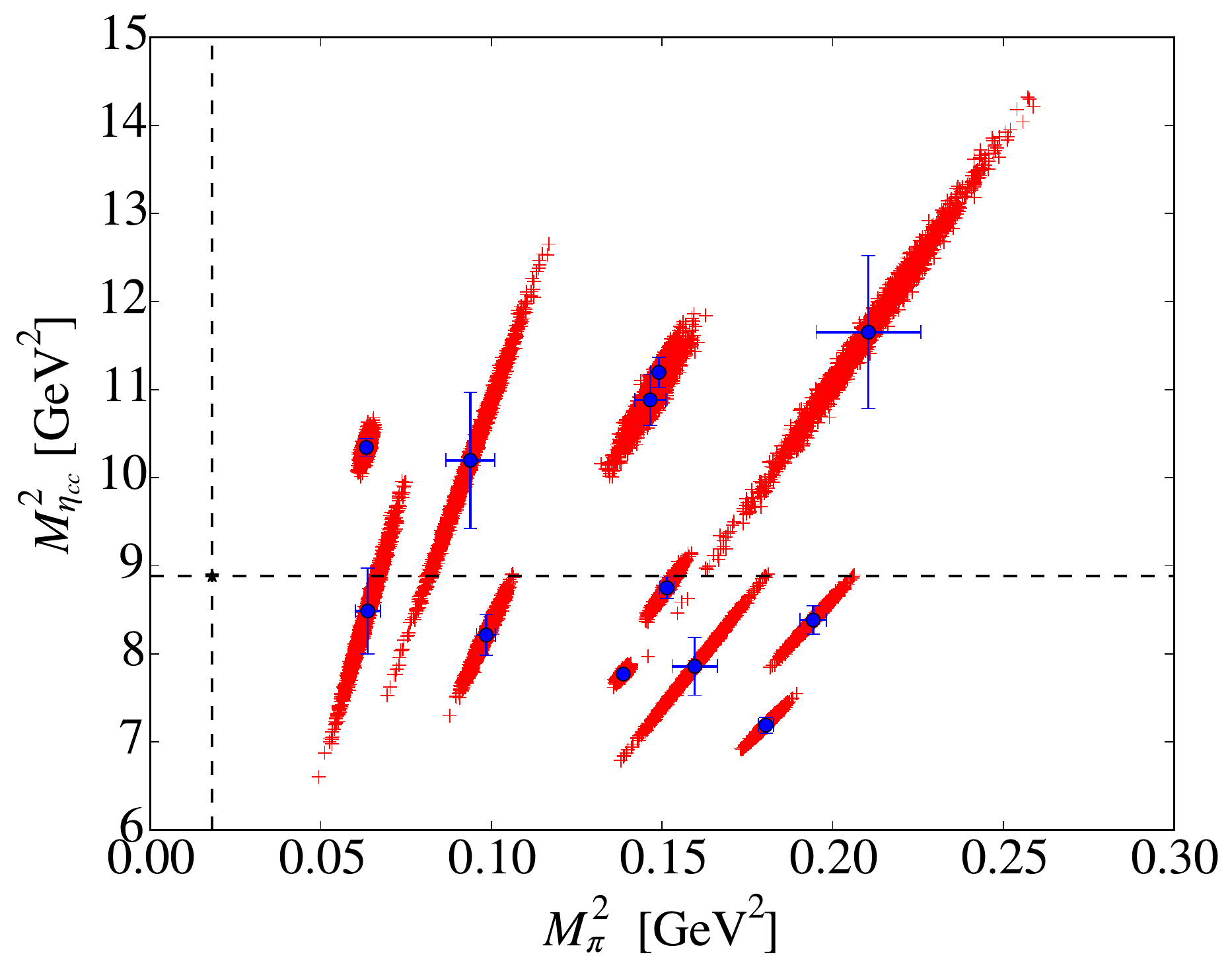}%
\includegraphics[width=0.5\textwidth,trim={8 10 5 5}, clip]{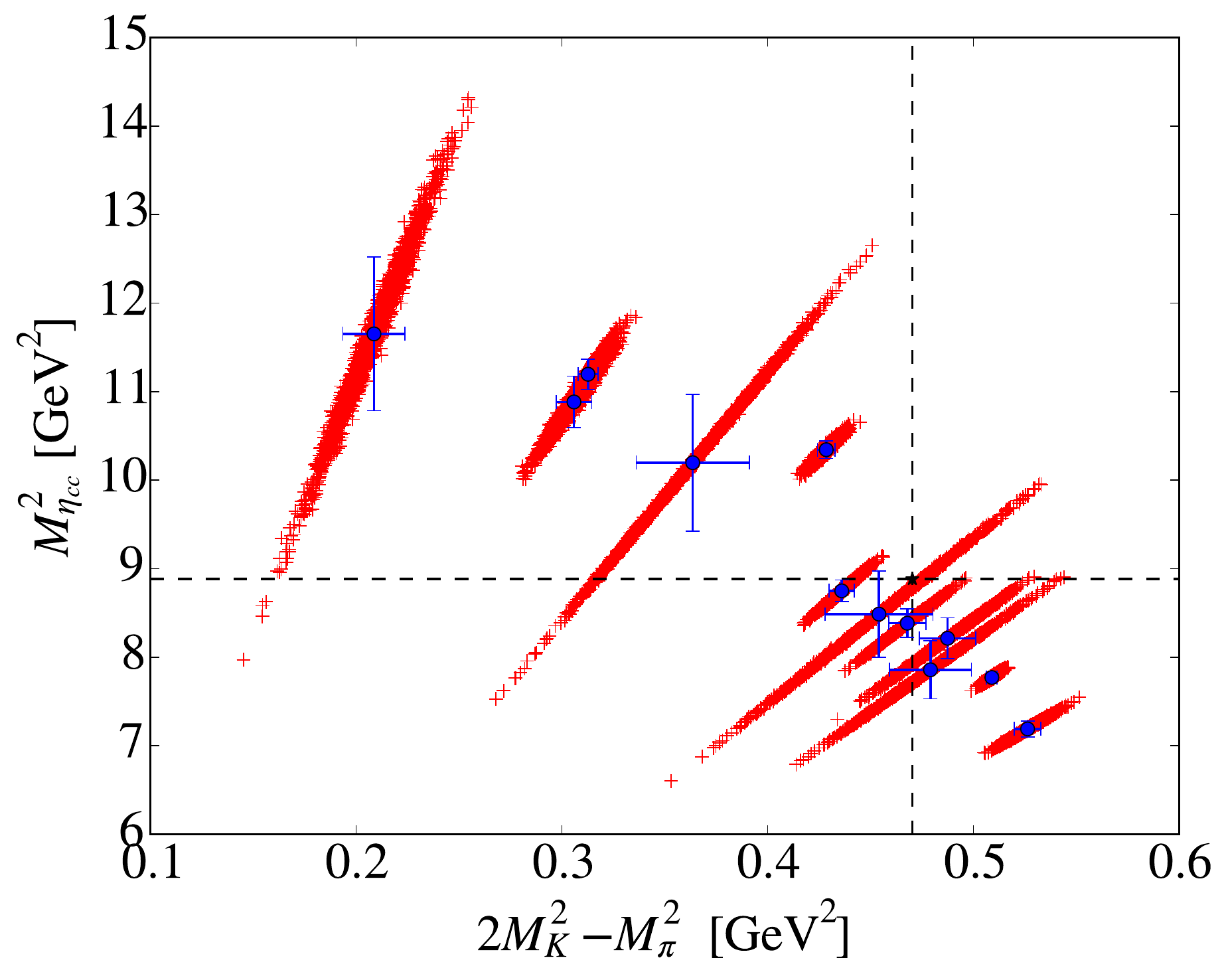}
\end{center}
\caption{Landscape plots for the $\beta=3.2$ ensembles. The red points indicate the individual resample averages and the blue points with error bars indicate the ensemble average with resample error. The black asterisk indicates the physical mass point. More ensemble details may be found in Table S3 of \cite{Borsanyi:2014jba}.}
\label{fig:lscape}
\end{figure}

\noindent For this calculation, we use the 27 `neutral' BMW ensembles \cite{Borsanyi:2014jba}, generated using the clover improved Wilson action with $N_f=1+1+1+1$ and 3 levels of HEX gauge link smearing \cite{Capitani:2006ni}. The ensembles are spread over 4 lattice spacings (Table \ref{ta:spa}) and have a pion mass range from 195 to 440 MeV (leaving a mild extrapolation in the light quark mass to the physical mass), whereas the strange and charm quark masses are chosen to be roughly physical. Our measurements are separated by 10 trajectories, with typically 400 configurations per ensemble and 64 source positions per configuration. The whole calculation is performed using a bootstrap analysis with 2000 resamples. To illustrate the distribution of our ensembles, landscape plots for the $\beta=3.2$ ensembles are shown in Fig.~\ref{fig:lscape}. One of the main challenges for charm physics calculations is that the large size of $am_c$ often leads to large cutoff effects due to the small correlation lengths (see Table \ref{ta:spa}).
\begin{table}[b]
\begin{center}
\begin{tabular}{|c|c|c|c|c|c|}
\hline
$\beta=6/g^2$ & $a[\text{fm}]$ & $am_c$ & $L^3\times T$ & $M_\pi~[\textrm{MeV}]$  min-max& no. ensembles \\
\hline
3.2 & 0.102 & 0.71 &  $32^3\times64$    & 235-440 & 12 \\
3.3 & 0.089 & 0.58 &  $32^3\times64$    & 195-410 & 6 \\
3.4 & 0.077 & 0.47 &  $32^3\times64$    & 220-405 & 3 \\
3.5 & 0.064 & 0.35 &  $32^3\times64$,  $64^3\times96$ & 200-420 & 6 \\
\hline
\end{tabular}
\end{center}
\caption{\label{ta:spa} Lattice spacing, charm quark mass, box size, pion mass range and number of ensembles at this $\beta$ value; all for the four gauge couplings. A complete table of parameters can be found in \cite{Borsanyi:2014jba}.}
\end{table}

In order to extract the ground state masses (and amplitudes), we use single state fits to the correlation functions, Eq.~\eqref{eq:simcor1}. When we have multiple channels that access the same observables, we perform the fits simultaneously. For example, for the pion:
\begin{align}\label{eq:simcor1}
\left|C_{PP}(t)\right| &= c^2\left(e^{-Mt}+e^{-M(T-t)}\right),\nonumber\\
\left|C_{PA_4}(t)\right| = \left|C_{A_4P}(t)\right| &= cd\left(e^{-Mt}+e^{-M(T-t)}\right),\\
\left|C_{A_4A_4}(t)\right| &= d^2\left(e^{-Mt}+e^{-M(T-t)}\right).\nonumber
\end{align}
The calculation was performed for both Gauss-Gauss (source-sink) and Gauss-point smearing, which totals 8 channels, and thus 5 fit parameters, $M$, $c$, $d$, $c'\rightarrow g_\pi$ and $d'\rightarrow f_\pi$, can be extracted (the prime indicates point amplitudes). It should be noted that, where appropriate, we average the measured observables in isospin multiplets, i.e. the 3 pions ($\pi^+$, $\pi^0$, $\pi^-$) and 2 kaons ($K^+$, $K^0$). 

Setting the lattice scale, so as to interpret things in physical units,  can be performed either globally or locally ensemble by ensemble (the choice for the results presented here). Both are equally valid and will feature in the systematic error budget of the final analysis (see Section \ref{sys}). For the normalisation, we use the omega mass, which we found could also be reliably extracted using one state fits to the correlation function.

\begin{figure}[h!]
\begin{center}
\includegraphics[width=0.5\textwidth, trim={20 10 130 30}, clip]{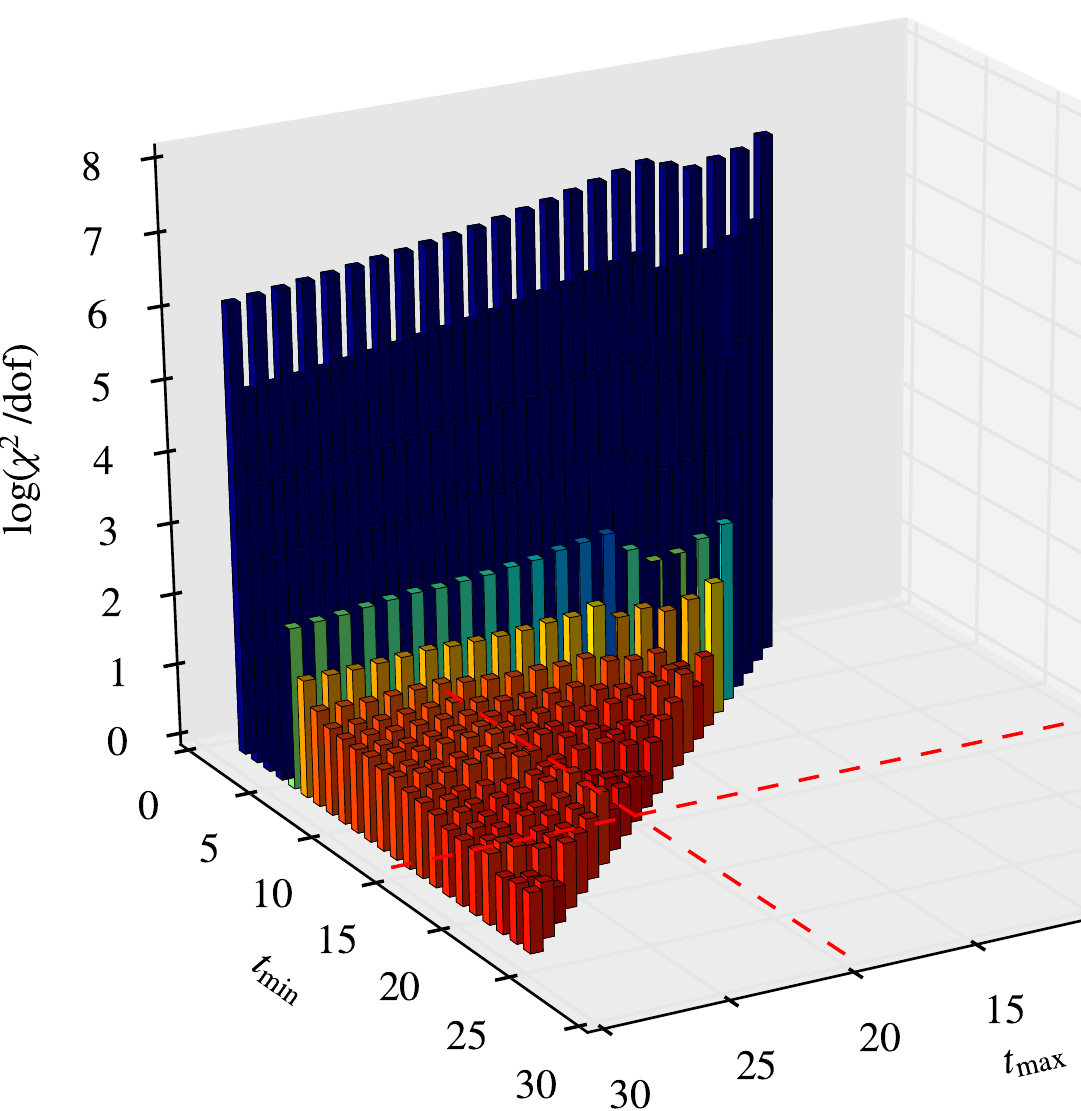}%
\includegraphics[width=0.5\textwidth, trim={20 10 130 30}, clip]{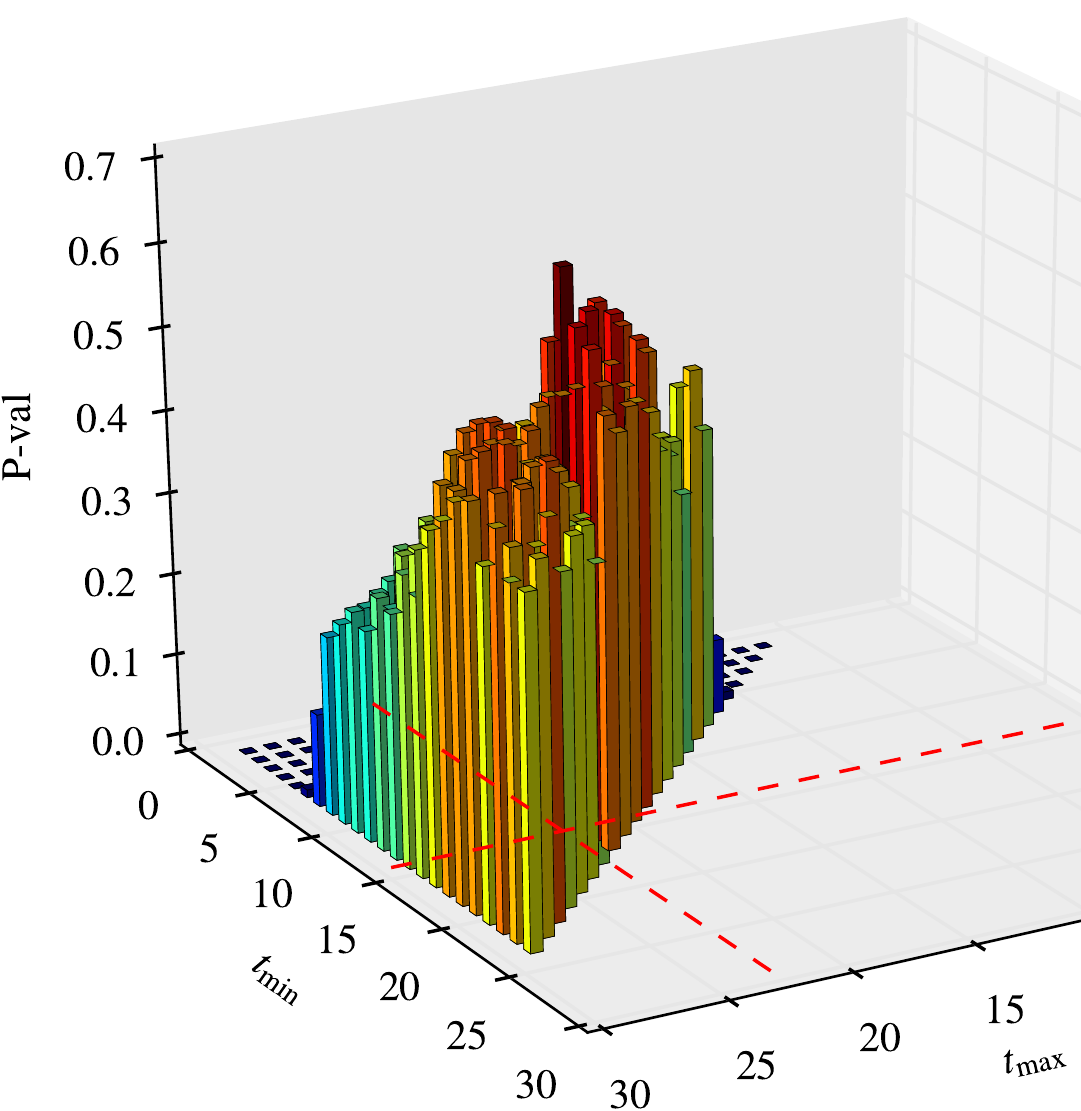}
\end{center}
\caption{Fitting interval selection criteria, $\chi^2/\textrm{dof}$ (left panel) and P-value (right panel). This demonstrates the selected fit interval $t_{\textrm{min}}$ to $t_{\textrm{max}}$ for a given channel for our $\beta=3.2$, $M_\pi=440$~MeV ensemble. The red dashed lines indicate the selected interval corresponding to the smallest ($\chi^2/\textrm{dof}$) and largest (P-value) bars.}
\label{fig:fitint}
\end{figure}

\begin{figure}[h!]
\begin{center}
\includegraphics[width=0.5\textwidth,trim={5 10 5 5}, clip]{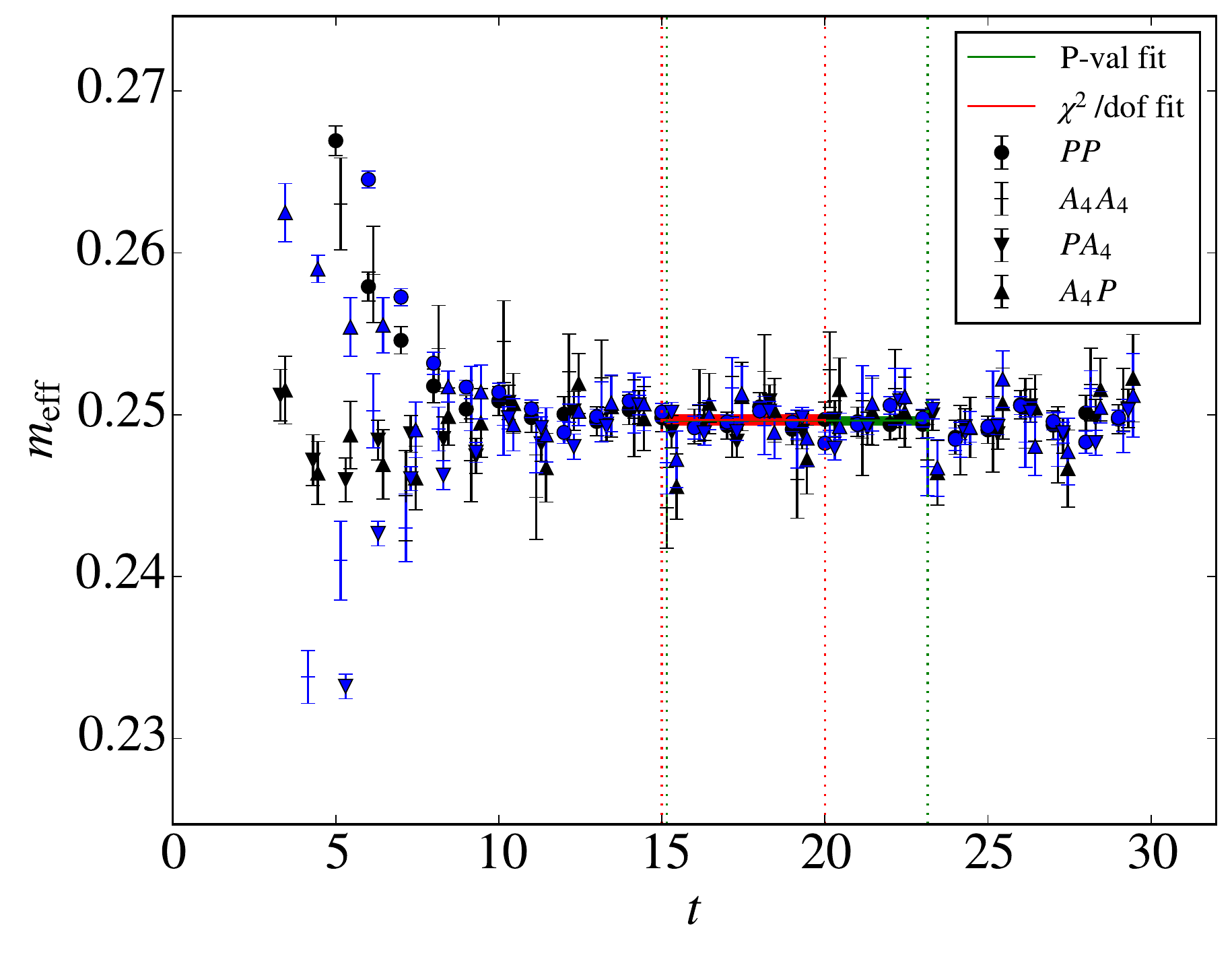}%
\includegraphics[width=0.5\textwidth,trim={5 10 5 5}, clip]{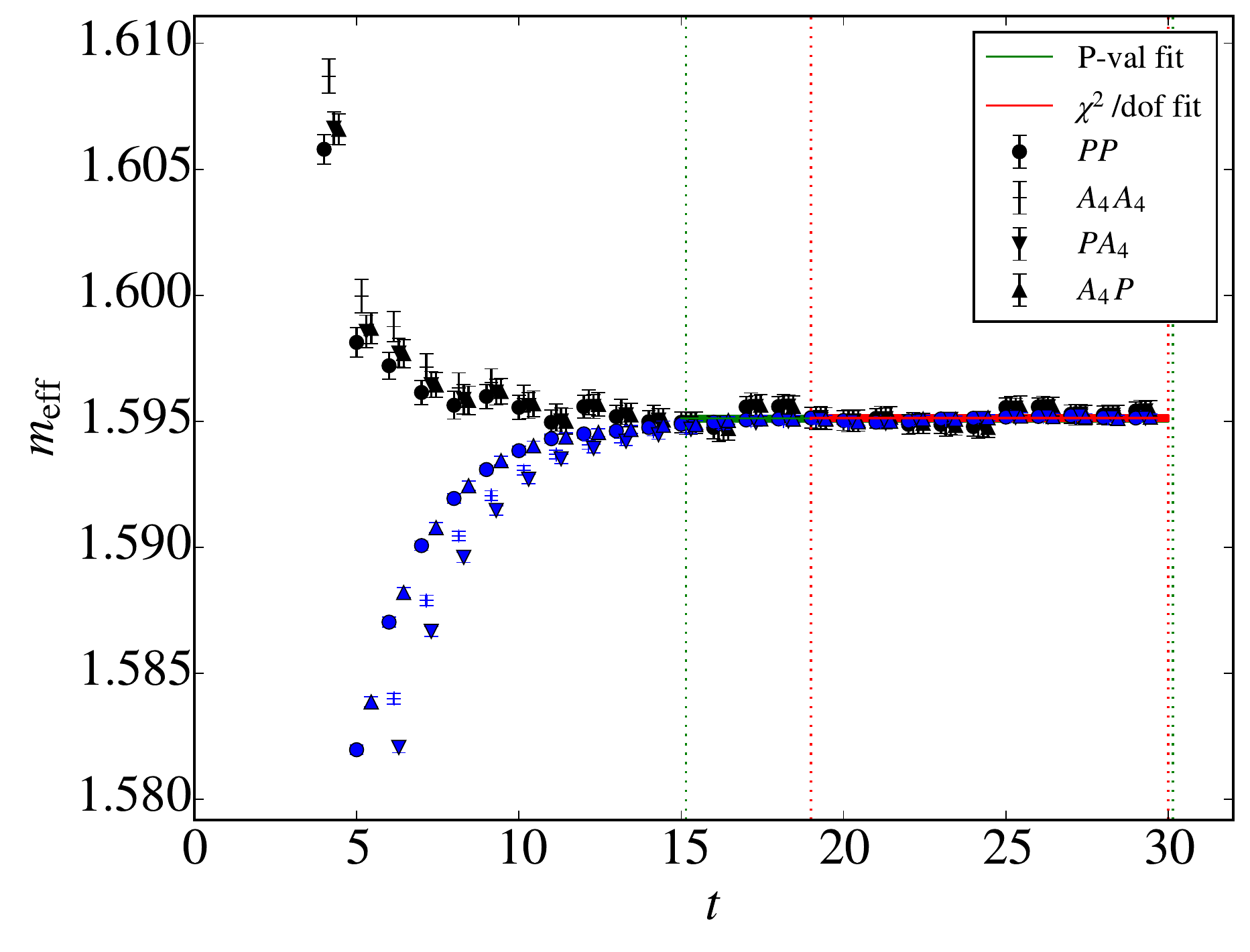}
\end{center}
\vspace{-0.5cm}
\caption{Effective mass plots for light-light and charm-charm channels (left and right panels respectively) for the $\beta=3.2$, $M_\pi=440$~MeV ensemble. The effective masses for the 8 channels, 4 Gauss-Gauss (source-sink) and 4 Gauss-point smearing, are shown by the black and blue points respectively. Point styles indicate the different channels, the coloured bars indicate the extracted mass and error from the simultaneous one state fits to the correlation functions and the vertical dotted lines indicate the automatically selected fitting interval.}
\label{fig:meff}
\end{figure}

One of the main challenges in fitting correlation functions is the determination of the fitting interval for any given channel. For this, we loop over all possible $t_\mathrm{min}$ and $t_\mathrm{max}$ intervals with the condition that there is a minimum separation of 5 timesteps. For each of these, we perform a fit using a correlated chi-squared minimisation, from which two fitting intervals are chosen for each channel; one using the smallest $\chi^2/\mathrm{dof}$ and one using the largest P-value. An example is shown in Fig.~\ref{fig:fitint}.
Given the selected fit range, we extract the ground-state mass and amplitudes (decay constants) for each of the 2000 bootstrap samples. Fig.~\ref{fig:meff} demonstrates the result for the mass extracted from the simultaneous one state fit by overlaying it onto the effective mass plots for the 8 channels.

\section{Hyperfine splitting}

\begin{figure}[h!]
\begin{center}
\includegraphics[width=0.5\textwidth,keepaspectratio=]{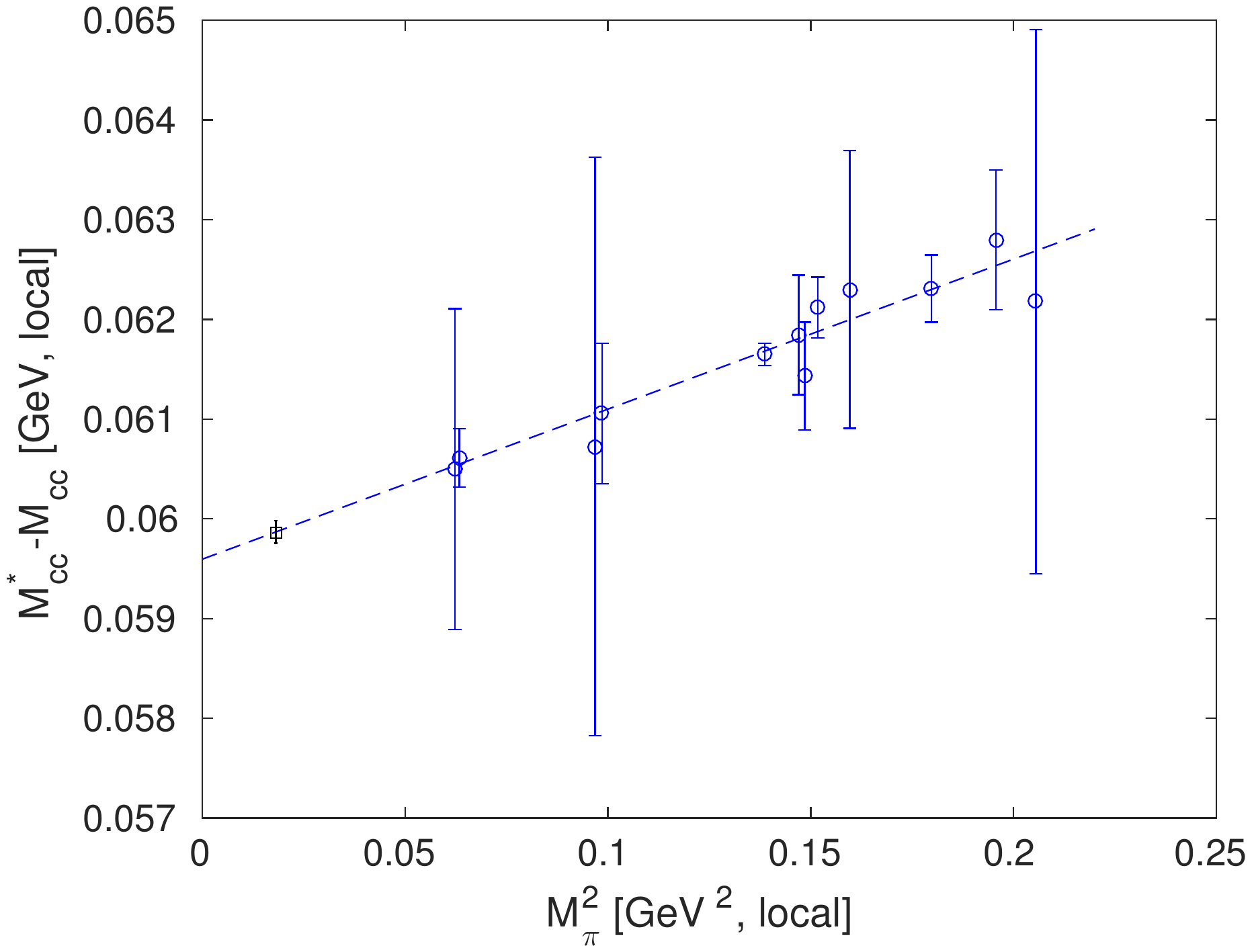}%
\includegraphics[width=0.5\textwidth,keepaspectratio=]{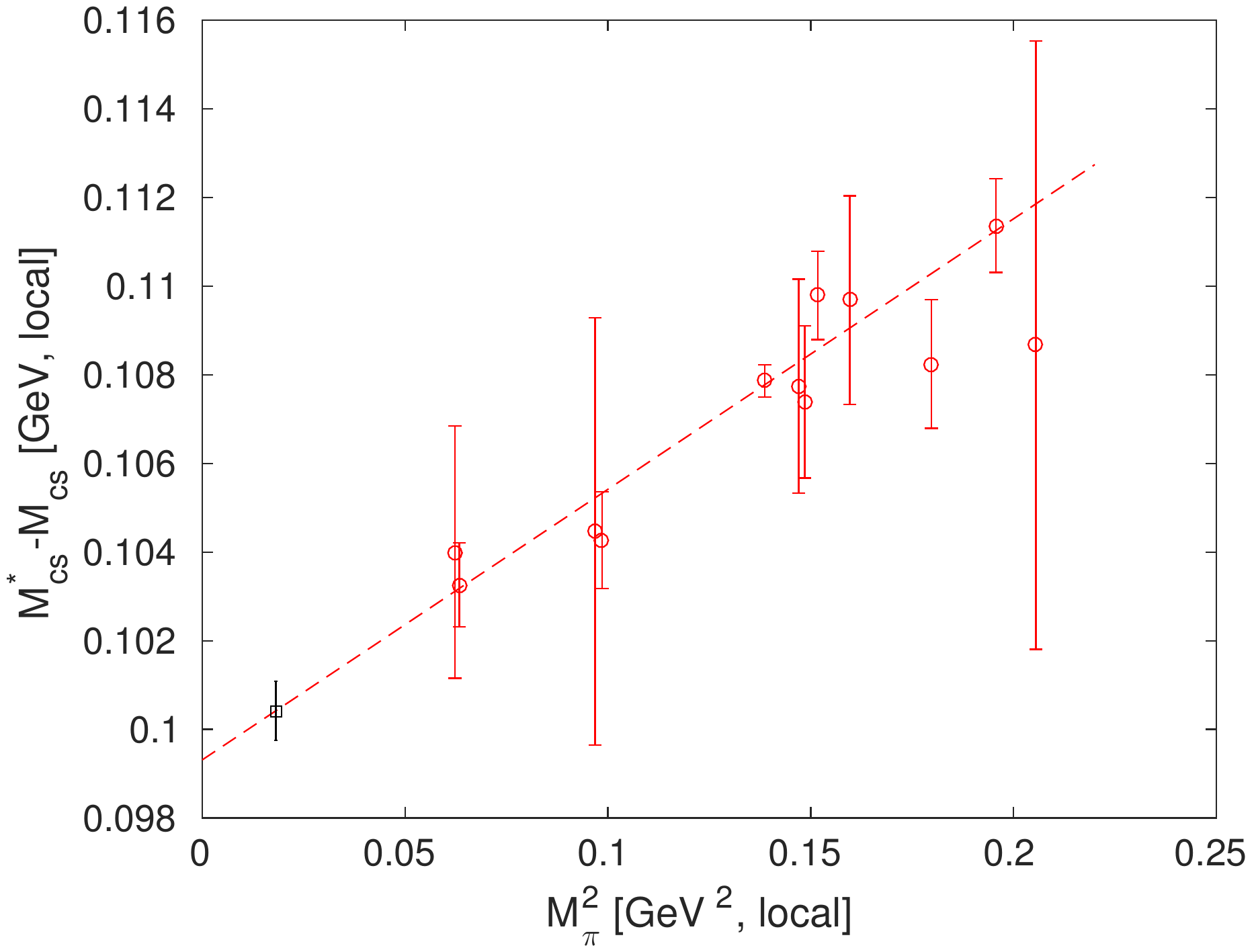}\\
\includegraphics[width=0.5\textwidth,keepaspectratio=]{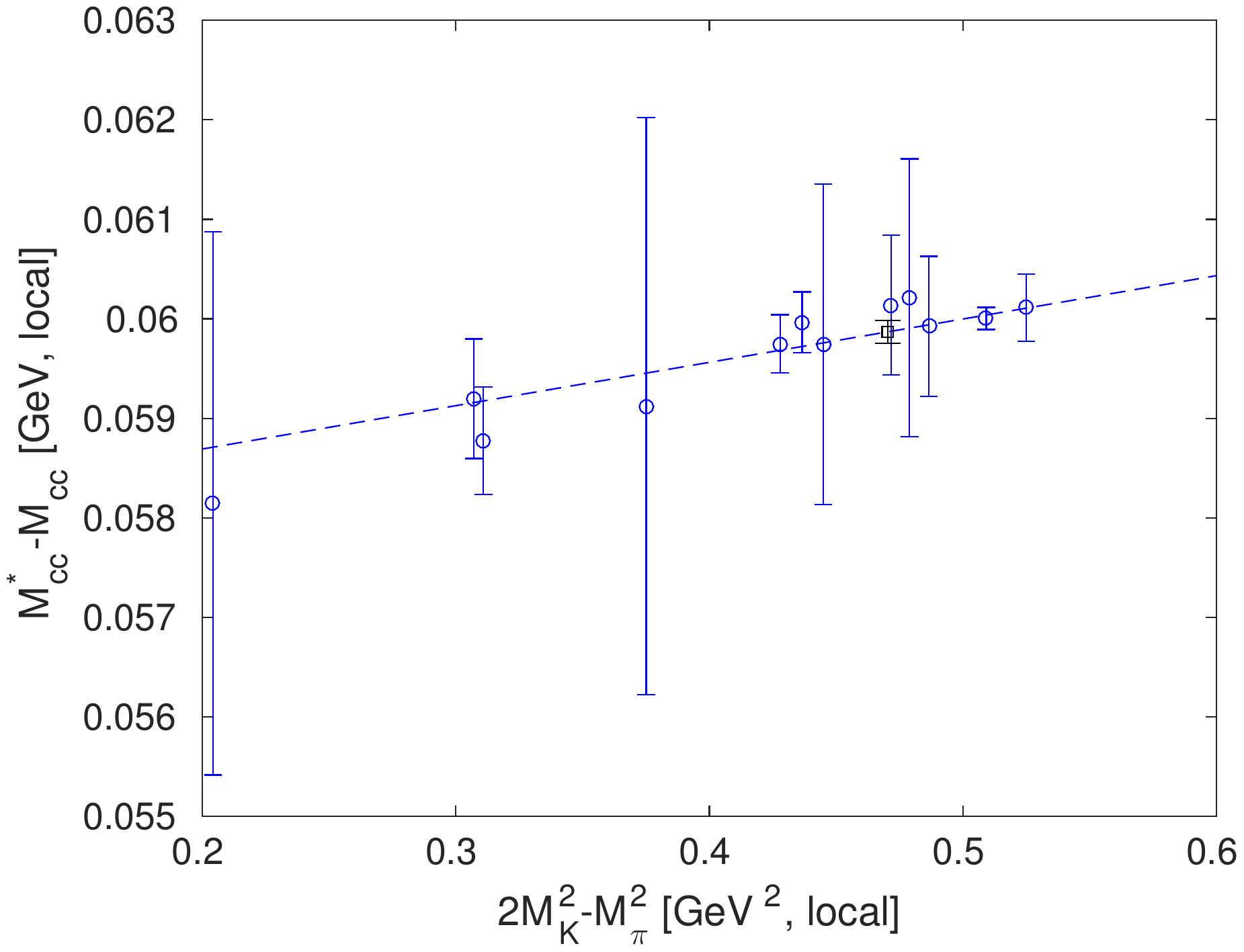}%
\includegraphics[width=0.5\textwidth,keepaspectratio=]{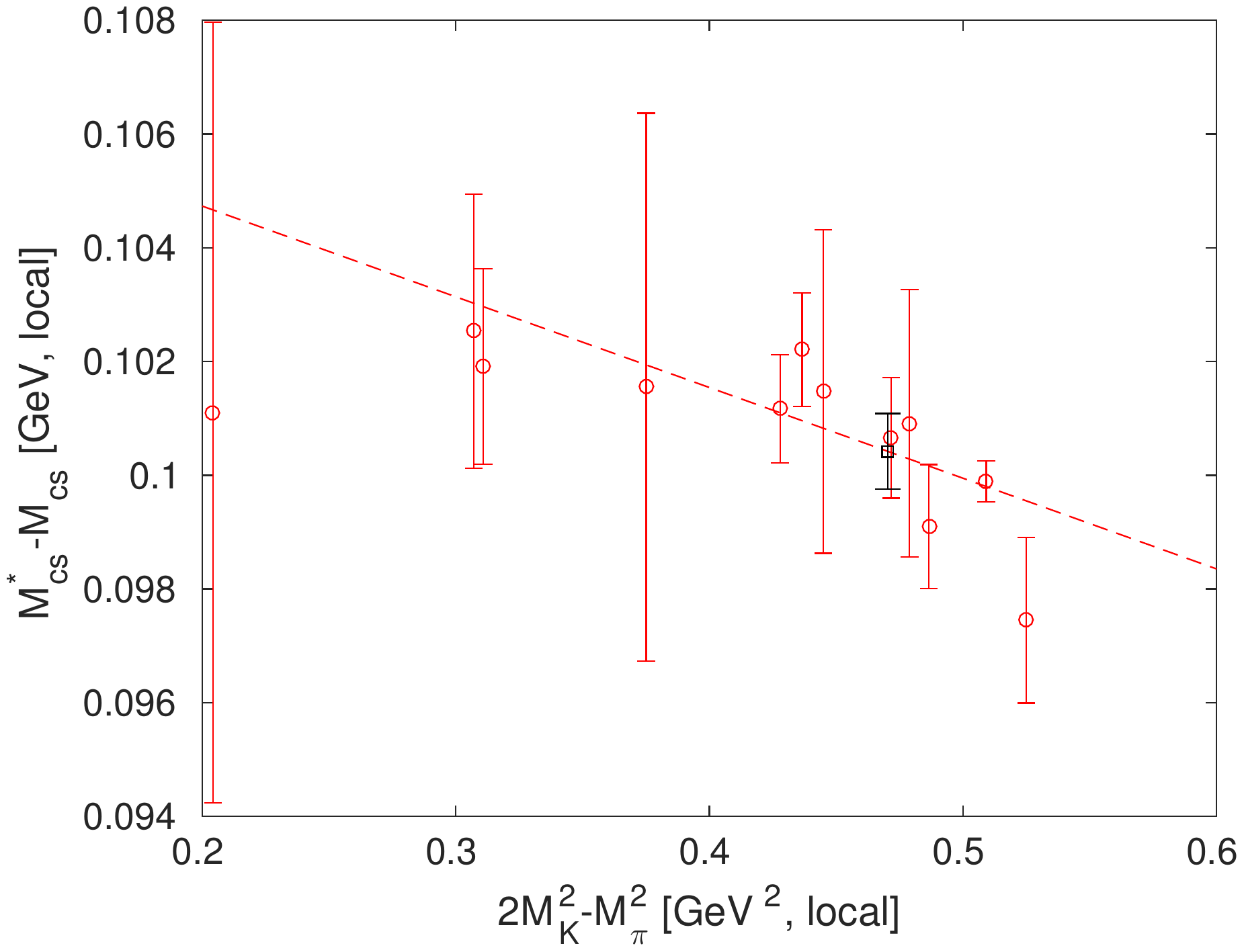}\\
\includegraphics[width=0.5\textwidth,keepaspectratio=]{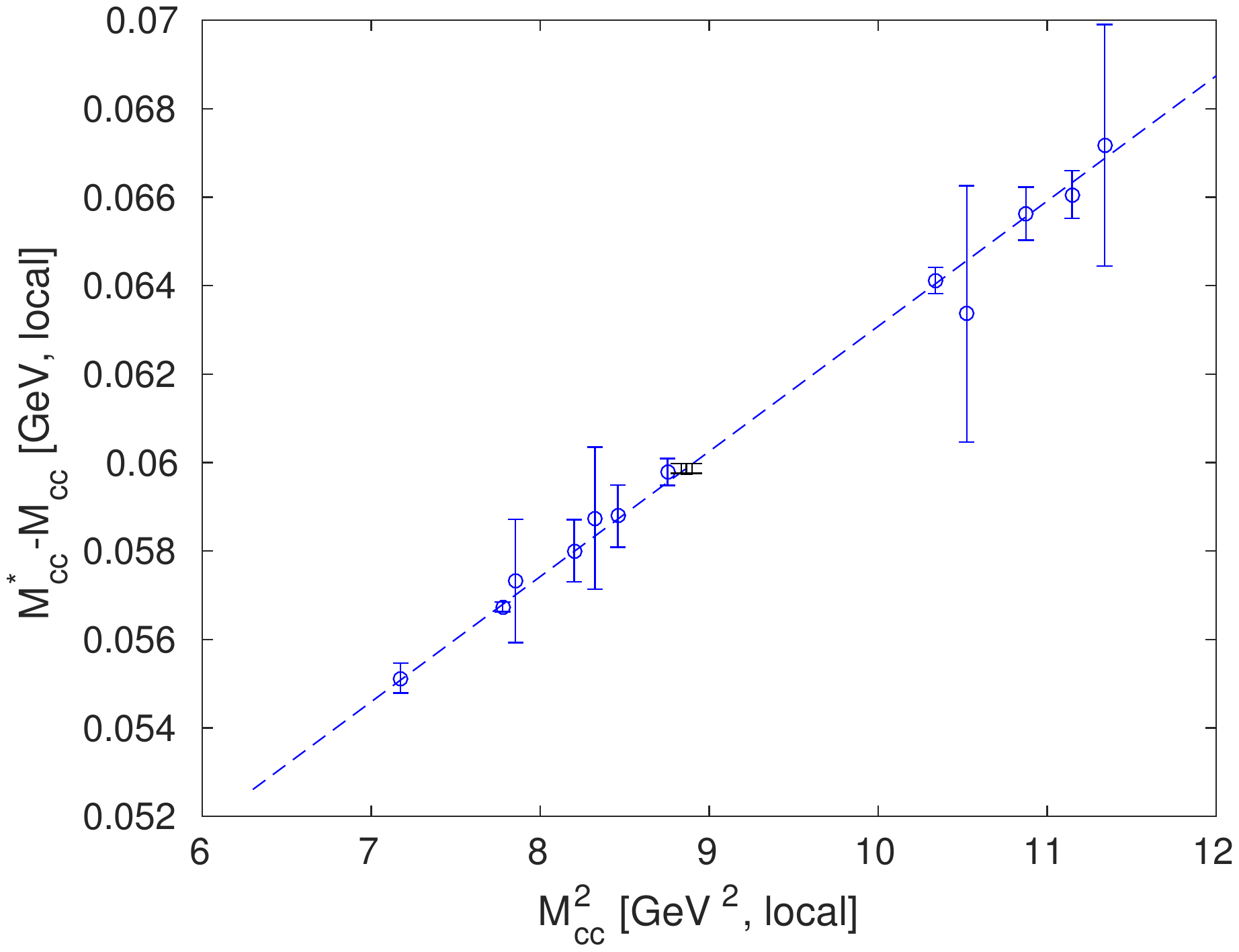}%
\includegraphics[width=0.5\textwidth,keepaspectratio=]{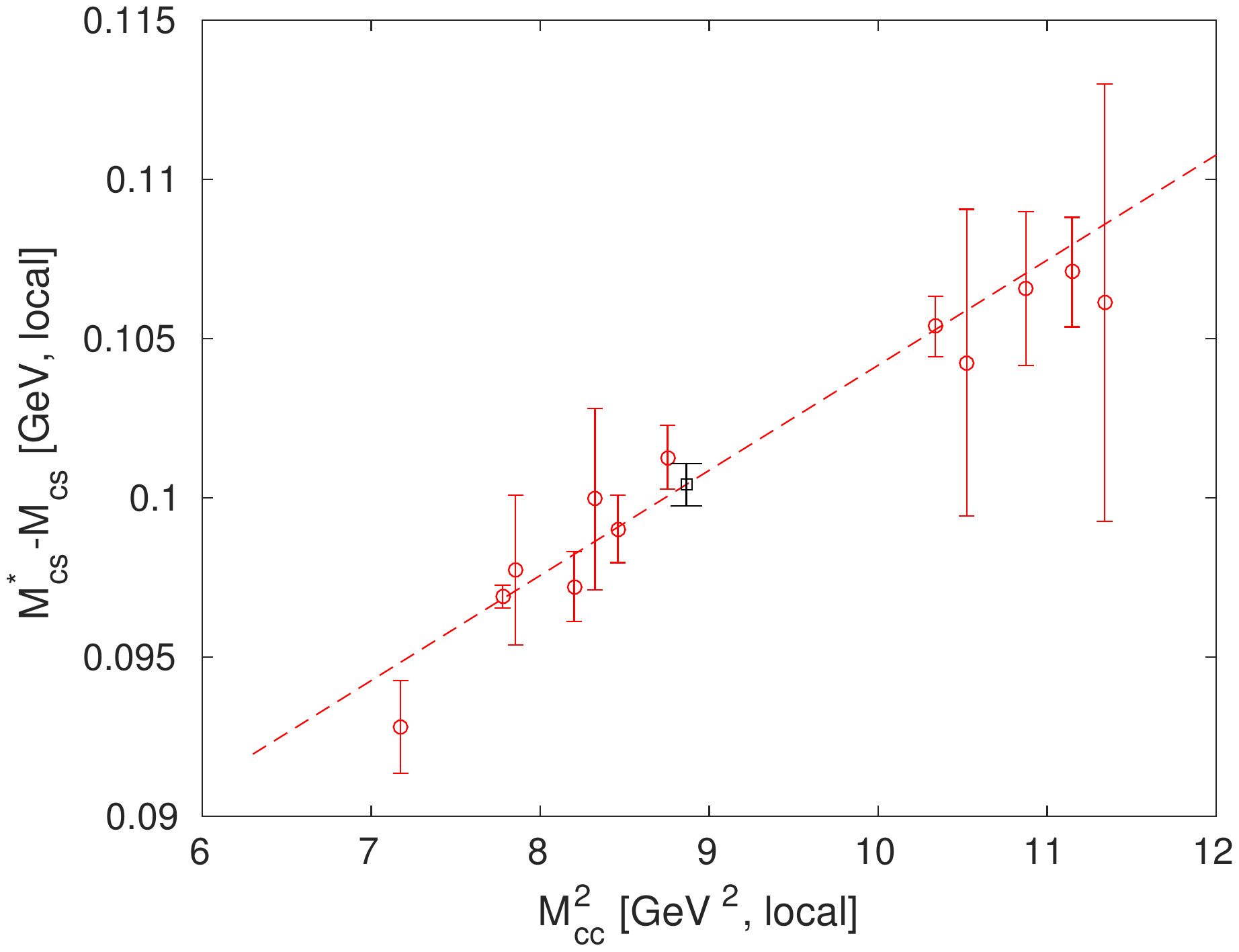}
\end{center}
\vspace{-0.5cm}
\caption{Results for the hyperfine splittings $M_{J/\psi}-M_{\eta_{cc}}$ (left panel) and $M_{D_s^*}-M_{D_s}$ (right panel), `projected' into two dimensional plots, where the x-axes are proxies for the light, strange and charm quark masses (top, middle and bottom rows respectively). The projection is such that the data are adjusted, by the fit, for the effects of the two coordinates that are not shown. The fitted value is given by the black square.}
\label{fig:ds}
\end{figure}
\begin{figure}
\begin{center}
\includegraphics[width=0.5\textwidth,keepaspectratio=]{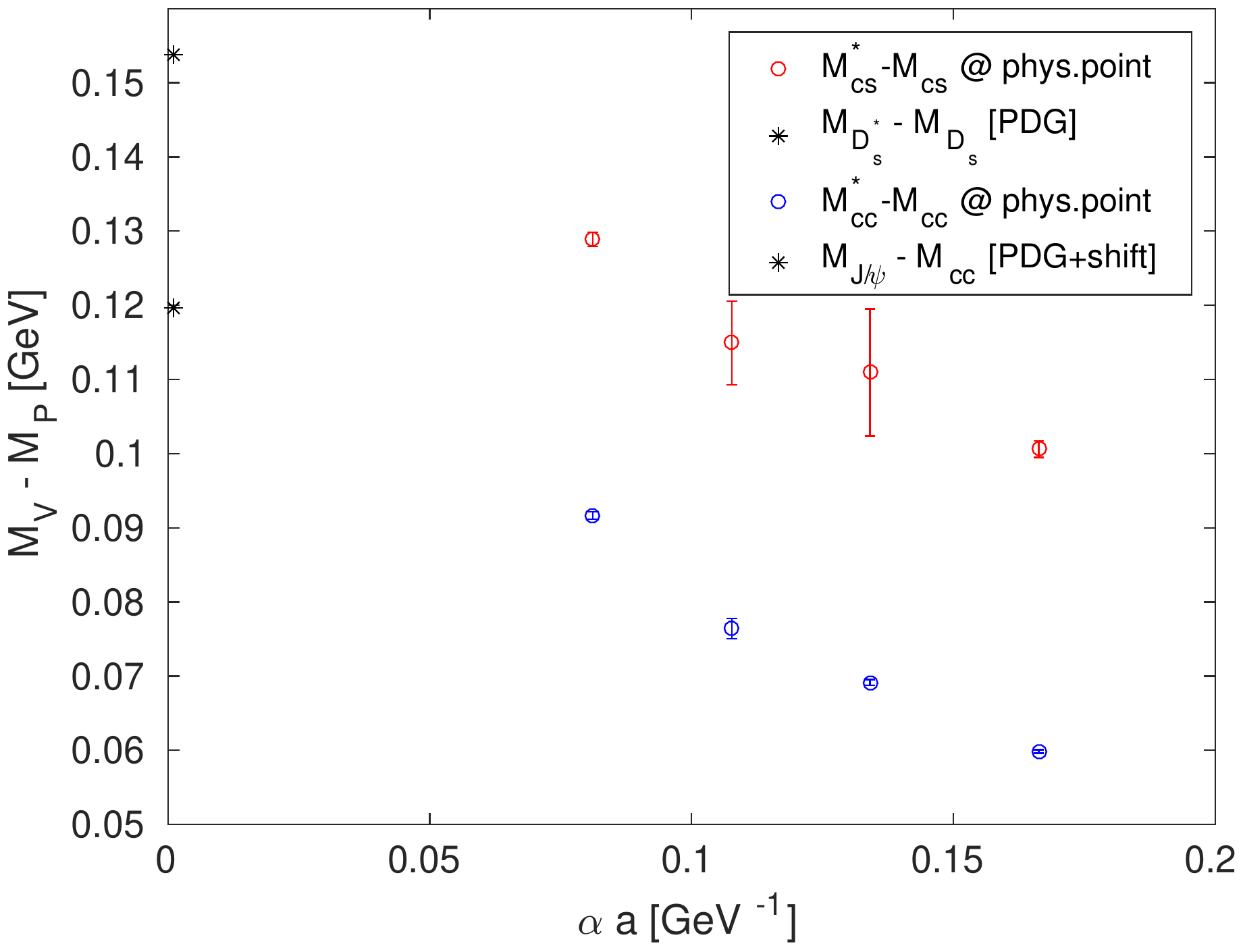}%
\includegraphics[width=0.5\textwidth,keepaspectratio=]{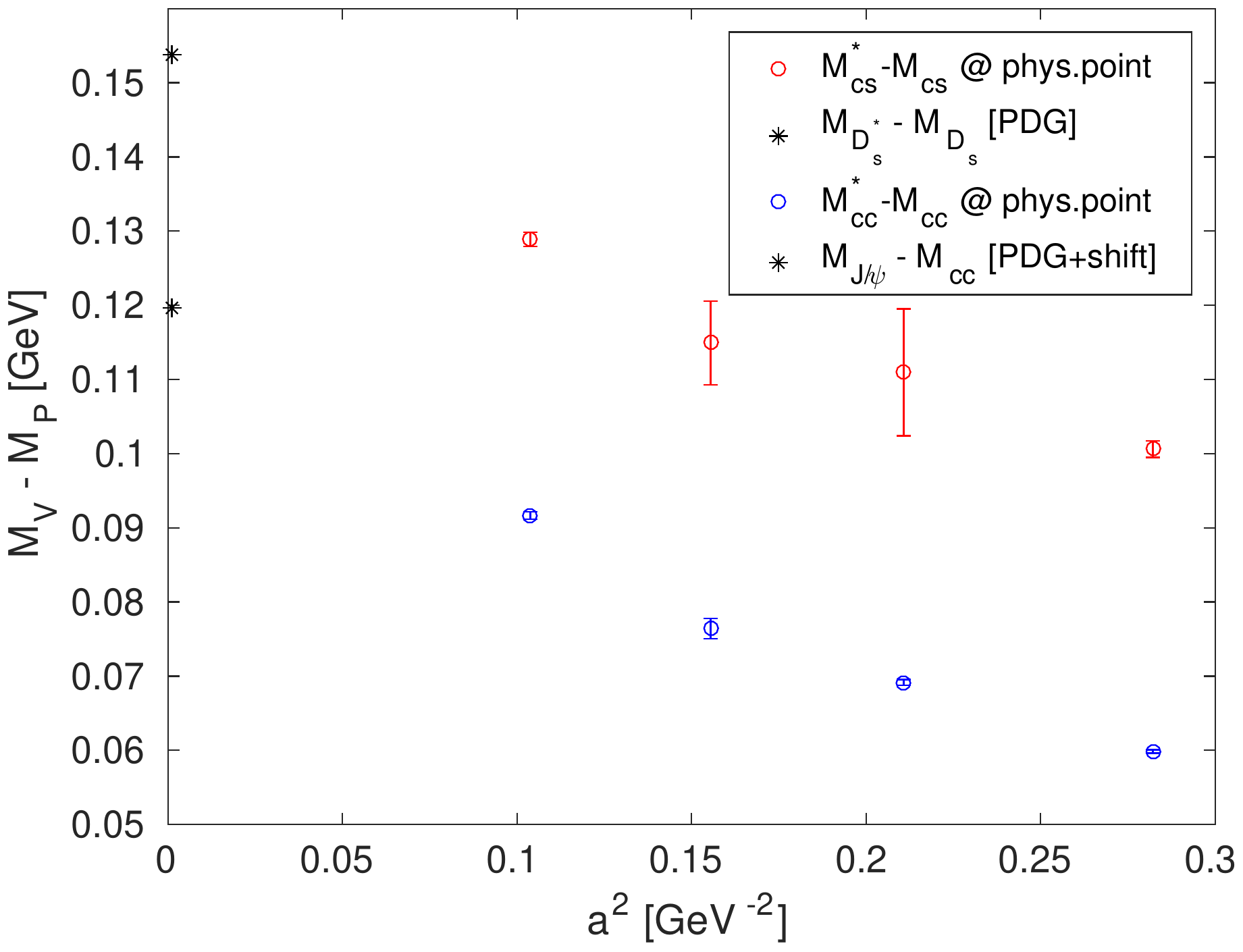}
\end{center}
\caption{Results for hyperfine splittings $M_{J/\psi}-M_{\eta_{cc}}$ (blue points) and $M_{D_s^*}-M_{D_s}$ (red points) versus $\alpha a$ (left panel) and $a^2$ (right panel). The errors are statistical only; a full analysis with a thorough treatment of all sources of systematic uncertainties is currently underway. The black asterisks show the PDG values. However it should be noted that the effects of quark-disconnected diagrams are not included in our calculation and therefore the PDG value for the $c\bar{c}$ splitting has been modified (for more details see text). 
}
\label{fig:latspace}
\end{figure}

\noindent The mass splittings of the doubly and singly charmed states that we are interested in are defined as:
\begin{equation}\label{eq:ansatz1}
f(x,y,z)=\frac{\left(M_{J/\psi}-M_{\eta_{cc}}\right)}{M_\Omega},
\qquad f(x,y,z)=\frac{\left(M_{D_s^*}-M_{D_s}\right)}{M_\Omega}
\end{equation}
where we are interested in the behaviour as a function of the light, strange and charm quark masses, for which we use the following proxies:
\begin{equation}\label{eq:xyz}
x = M_\pi^2/M_\Omega^2,\quad
y = \left(2M_K^2-M_\pi^2\right)/M_\Omega^2,\quad
z = M_{\eta_{cc}}^2/M_\Omega^2.
\end{equation}
We found that a good description of the chiral behaviour, for both splittings, could be obtained using the linear ansatz
\begin{equation}\label{eq:ansatz}
f(x,y,z)=c_0+c_1 x+ c_2 y + c_3 z,
\end{equation}
which is shown as a two dimensional projection in Fig.~\ref{fig:ds} (see caption for details). We can see the mild extrapolation to the physical point in the light quark mass and the interpolation to the physical point in both the charm and strange quark masses. This is shown for the $\beta=3.2$ ensembles, and is repeated for each individual beta so that we may study the scaling window when extrapolating to the continuum limit (Fig.~\ref{fig:latspace}). We are also investigating alternative ans\"atze for the chiral behaviour.

For the continuum extrapolation, we look at the behaviour in $\calO(\alpha_s a)$ and $\calO(a^2)$, i.e. the leading discretisation errors for our particular lattice action (see \cite{Borsanyi:2014jba} for details). Fig.~\ref{fig:latspace} shows a significant dependence of the mass splitting on the lattice spacing, however the lattice measurements appear consistent with the physical limit. 
The effects of quark disconnected diagrams are not included in our simulations. Without these the $c\bar{c}$ hyperfine splitting increases by a few MeV \cite{Feldmann:1998sh,Cheng:2008ss}. This has been estimated from LQCD simulations \cite{Levkova:2010ft} to be $\sim3$~MeV\footnote{It should be noted that in \cite{Follana:2006rc}, a calculation using perturbation theory predicts a contribution with the opposite sign, thus decreasing the hyperfine splitting.}. To reflect this, we have shifted the PDG value for the $c\bar{c}$ mass splitting in Fig.~\ref{fig:latspace} by $\sim3$~MeV.

It should be noted that Fig.~\ref{fig:latspace} shows only the statistical errors and should be considered a `snapshot' plot, i.e. it is only for one particular choice of parameters (see Section \ref{sys}) and, as such, does not include any systematic errors. A full systematic analysis is ongoing, the considerations of which are detailed in the next section. Having checked the scaling window, we will proceed with a thorough systematic analysis using a `global' fitting procedure, that simultaneously fits all betas through the addition of a term proportional to $\alpha_s a$ or $a^2$.

\section{Systematic analysis and outlook}\label{sys}

\noindent As with all lattice calculations, at various points throughout the calculation several, equally valid, choices can be made (listed below), and it is through performing all of these on a level footing that we can give a thorough and detailed systematic error budget. All possible combinations of the following will be evaluated within a looped bootstrap analysis:
\begin{itemize}
\item Fitting interval selection: $\chi^2/$dof versus P-value (Fig.~\ref{fig:fitint} left versus right panels).
\item Choice of scale setting procedure: local versus global.
\item Chiral ansatz: Eq.~\ref{eq:ansatz} versus alternatives.
\item Continuum extrapolation: $\alpha_s a$ versus $a^2$ (Fig.~\ref{fig:latspace} left versus right panels).
\item Inclusion of finite volume effects through the comparison of different ans\"atze.
\end{itemize}

The extracted data also allows us to look at the charmed decay constants, which will be treated analogously to the mass splittings and will appear in a forthcoming paper.

{~\\ \noindent {\bf Acknowledgments:}
We thank Keh-Fei Liu for useful discussions regarding the disconnected mass shift of the $\eta_c$. This project was supported by the Deutsche Forschungsgemeinschaft grant SFB/TR55. The computations were performed on JUQUEEN and JUROPA at Forschungszentrum J\"ulich (FZJ), on Turing at IDRIS in Orsay, on SuperMUC at Leibniz Supercomputing Centre in M\"unchen and on Hermit at the High Performance Computing Center in Stuttgart.
}


\begin{thebibliography}{99}

\bibitem{Brambilla:2010cs}
  N.~Brambilla {\it et al.},
  Eur.\ Phys.\ J.\ C {\bf 71} (2011) 1534
  [arXiv:1010.5827 [hep-ph]].
  
\bibitem{Amhis:2014hma}
  Y.~Amhis {\it et al.} [Heavy Flavor Averaging Group (HFAG) Collaboration],
  arXiv:1412.7515 [hep-ex].

\bibitem{Bouchard:2015pda}
  C.~M.~Bouchard,
  PoS LATTICE {\bf 2014} (2015) 002
  [arXiv:1501.03204 [hep-lat]].
  
\bibitem{El-Khadra:2014sha}
  A.~X.~El-Khadra,
  PoS LATTICE {\bf 2013} (2014) 001
  [arXiv:1403.5252 [hep-lat]].
  
\bibitem{Aoki:2013ldr}
  S.~Aoki {\it et al.},
  Eur.\ Phys.\ J.\ C {\bf 74} (2014) 2890
  [arXiv:1310.8555 [hep-lat]].

\bibitem{Bali:2011dc}
  G.~Bali {\it et al.},
  PoS LATTICE {\bf 2011} (2011) 135
  [arXiv:1108.6147 [hep-lat]].

\bibitem{Borsanyi:2014jba}
  S.~Borsanyi {\it et al.},
  Science {\bf 347} (2015) 1452
  [arXiv:1406.4088 [hep-lat]].
  
\bibitem{Capitani:2006ni}
  S.~Capitani, S.~Durr and C.~Hoelbling,
  JHEP {\bf 0611} (2006) 028
  [hep-lat/0607006].

\bibitem{Feldmann:1998sh}
  T.~Feldmann, P.~Kroll and B.~Stech,
  Phys.\ Lett.\ B {\bf 449} (1999) 339
  [hep-ph/9812269].
  
\bibitem{Cheng:2008ss}
  H.~Y.~Cheng, H.~n.~Li and K.~F.~Liu,
  Phys.\ Rev.\ D {\bf 79} (2009) 014024
  [arXiv:0811.2577 [hep-ph]].

\bibitem{Levkova:2010ft}
  L.~Levkova and C.~DeTar,
  Phys.\ Rev.\ D {\bf 83} (2011) 074504
  [arXiv:1012.1837 [hep-lat]].
  
\bibitem{Follana:2006rc}
  E.~Follana {\it et al.} [HPQCD and UKQCD Collaborations],
  Phys.\ Rev.\ D {\bf 75} (2007) 054502
  [hep-lat/0610092].

\end{thebibliography}
\end{document}